# Strong-Field Photoemission Experiment using Diamond Field Emitter Arrays

Vitaly Pavlenko[1, a)], Dongsung Kim[1, b)], Heather L. Andrews[1], Dmitry V. Gorelov[1], Evgenya I. Simakov[1]

[1]*Los Alamos National Laboratory, AOT-AE, Los Alamos National, USA*
a)pavlenko@lanl.gov, b)dskim84@lanl.gov

**Abstract.** In this paper, we present the results of experimental observation of strong-field photoemission from a diamond field-emitter array (DFEA) illuminated by a focused laser beam with 1035 nm wavelength. Having the advantage of high emission current and low beam emittance, DFEAs can emit ultra-short high charge electron bunches required for multiple accelerator applications. We have performed strong-field photoemission experiment at a newly commissioned test stand at Los Alamos National Laboratory (LANL). We triggered a diamond tip with a typical apex of 10-20 nm with 300 femtosecond (fs) laser pulses to produce electron beam emission. The profile of the emitted beam was determined by the analysis of an image formed on the fluorescent screen after a microchannel plate (MCP) intensifier; beamlets corresponding to emission from the pyramid sides and nano-tip were identified. Image analysis of the beamlet associated with emission from the nano-tip was conducted to understand dependencies of intensities of the emission on polarization of the laser's light and its peak intensity.

## INTRODUCTION

Needle cathodes are essential sources that can generate high quality electron beams for electron spectroscopy and other applications [1-3]. It was proposed to use these cathodes as laser-triggered electron emitters in dielectric laser accelerators (DLAs) and ultrafast electron microscopy [4-6], which require tightly focused ultra-short high current electron bunches. At Los Alamos National Laboratory (LANL) we proposed to use diamond field emitter array (DFEA) cathodes as an alternative to needle cathodes. DFEAs are known to deliver high per-tip current with low emittance via field emission [7]. Diamond has high thermal conductivity, is chemically inert, and performs well in poor vacuum conditions [7].

At LANL, we have established a procedure for fabricating DFEA cathodes and manufactured multiple DFEA samples that have been used in various experiments, for example the beam shaping experiment at Argonne National Laboratory [8-10]. Diamond cathodes have exquisitely sharp tips (nano-tips) on top of micrometer scale diamond pyramids. Using a mold-transfer fabrication process, we can fabricate DFEAs with various array configurations. Details of the fabrication process and a structural characterization of the arrays are reported elsewhere [11]. Although the modeling of field emission from nano-crystalline and diamond nano-tips is complicated [12], DFEAs are known to emit electron beams even under relatively low electric fields, and we conducted multiple experiments to demonstrate emission and characterize emitted electron beams in direct current (DC) and radio-frequency (rf) gun regimes [9, 10, 13]. Some applications such as DLAs require accurate synchronization of the electron emission with, e.g., accelerating laser pulses, so electron emission mechanism that can be triggered by an incident laser light is strongly desirable because generating ultrashort pulses with field emission is almost impossible. Photoemission mechanism unfortunately produces very strong emission from the substrate that dwarfs the emission from the tip, even if it is somewhat increased due to field enhancement [14]. We have previously demonstrated that the strong-field photoemission mechanism can be used to generate highly focused electron bunches from the diamond nano-tips [15] with efficiency that exceeds that of metallic needle cathodes [4]. Recently we assembled a test stand to study strong-field photoemission from a diamond tip triggered by intense laser pulses at 1035 nm.

Scanning electron microscope images of the DFEA cathode that was tested in this particular experiment are shown in Fig. 1. The cathode is a diamond square with 5x5 array of pyramids [Fig. 1(a)]. Each pyramid had a 13 μm base, and the spacing between pyramids was 400 μm [Fig. 1(a, b)]. There was an ultra-sharp tip (with diameter of 10-20 nm) on the top of each pyramid [Fig. 1(c)].

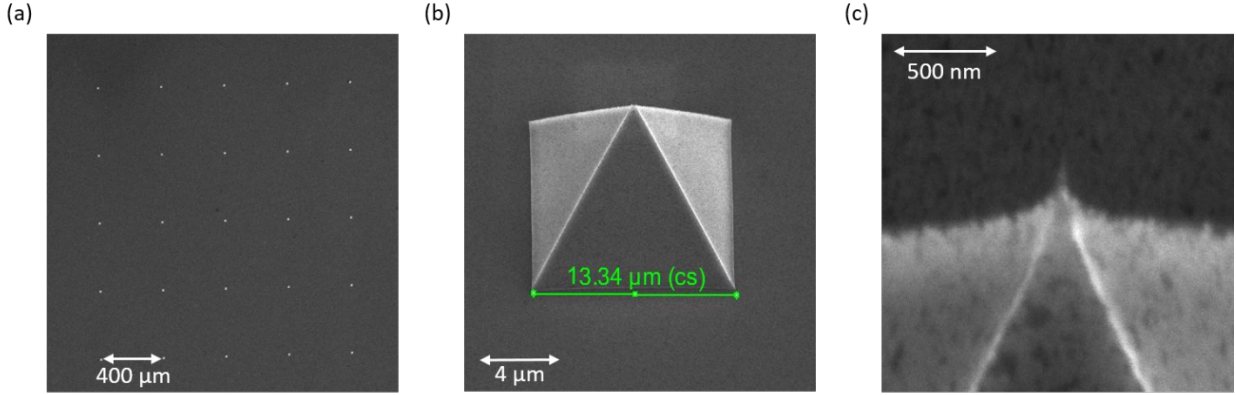

**FIGURE 1.** Scanning electron micrographs of the DFEA cathode. (a) 5x5 sparse array of diamond pyramids with 400 μm spacing, (b) a single diamond pyramid with a 13 μm base, (c) an ultra-sharp tip on the top of a pyramid.

## DETAILED EXPERIMENTAL SETUP

Figure 2 illustrates the setup of the strong-field photoemission experiment. The relevant parameters of the test stand are summarized in Table 1. The diamond cathode was mounted on a movable stage in an ultrahigh vacuum chamber. The position of the stage could be adjusted in Z-direction to change the anode-cathode gap (AK gap). The anode with a 1.5 mm aperture was mounted on a separate stage that could move in the transverse (XY) direction to precisely align the nano-tip with the aperture. Experiments were typically conducted at 10 kV voltage applied over 3 mm AK gap that was below the threshold for the DC field emission. By adjusting XYZ position of a parabolic mirror with a 152.4 mm focal length with respect to the sample, the focus of the laser light was approximately aligned on the pyramid with an aid of a camera. Fine tuning of laser focus was performed to obtain the strongest emission from the nano-tip as indicated by the brightness of the spot on the fluorescent screen. The measured beam spot diameter at the sample's position was about 100 μm. The average power of the laser was varied between 7.6 mW and 60.8 mW. Laser's repetition rate was varied between 100 kHz and 1 MHz with a stable 300 femtosecond (fs) pulse duration. The laser was directed on the sample at an angle of incidence of 6° to the substrate. After passing through the anode's aperture, the electrons are assumed to travel ballistically until they enter the microchannel plate (MCP) intensifier located approximately 30 cm from the anode. The emission patterns on the fluorescent screen after the MCP were captured by an optical camera under different condition of laser polarization and the power intensity. At any given fluorescent screen potential and voltage across the MCP stack (gain) the brightness of the digital images generated by the camera is proportional to the spatially resolved electron beam current density because the components of the system are essentially linear. Therefore, brightness of all pixels integrated over a certain region of interest (beamlet) can be interpreted as proportional to the total current of a beamlet. Accurate calibration of the image brightness versus electron beam current was not attempted, hence beam intensity reported in this work remains in arbitrary units, which is sufficient for qualitative interpretation of the results.

**TABLE 1.** Parameters for the strong-field photoemission experiment

| Parameters | Value |
|---|---|
| Base size of the pyramids | 13 μm |
| Spacing between pyramids | 400 μm |
| Voltage between the cathode and the anode | 10 kV |
| Anode-cathode gap | 3.0 mm |
| Diameter of the anode's aperture | 1.5 mm |
| Laser's wavelength | 1035 nm |
| Laser's pulse length | 300 fs |
| Laser's repetition rate | 100 kHz – 1 MHz |
| Laser's spot size (diameter) on the cathode | 100 μm |
| Laser's incident angle | 6° |

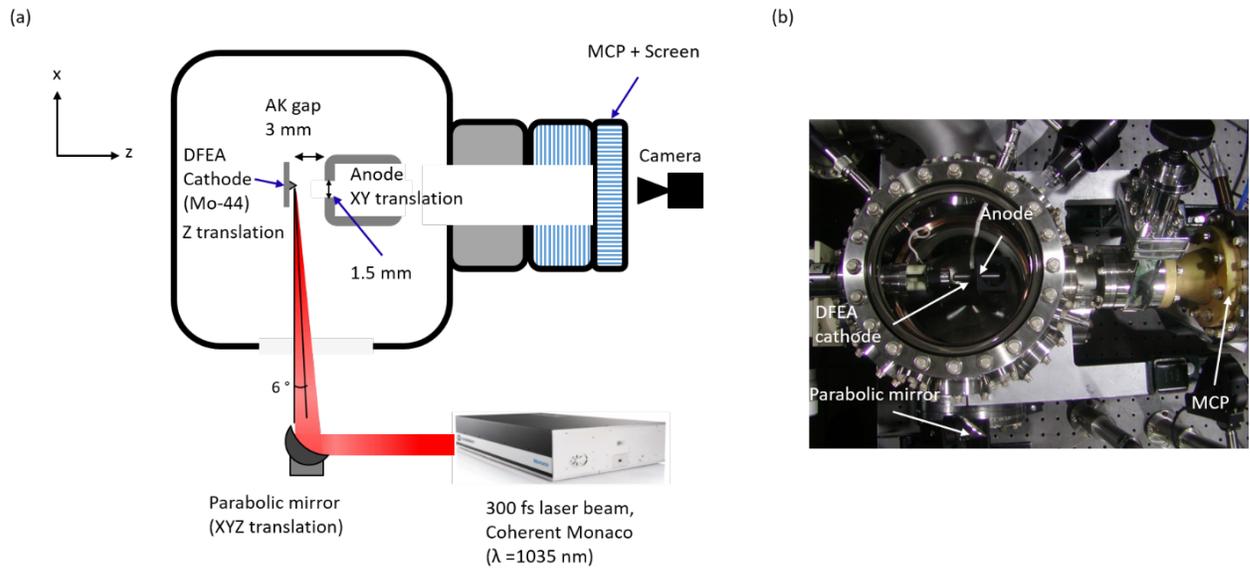

**FIGURE 2.** Experimental setup for strong-field photoemission. (a) The schematic of the experiment. (b) A photograph of the vacuum chamber.

## EXPERIMENTAL RESULTS

### 1. Measurement of the dependence of emitted current on polarization

As a first experiment, we conducted measurements of dependence of the emitted electron beam current on the angle of laser light polarization with respect to the pyramid tip. In this experiment, the output power and repetition rate of the laser were fixed at 14.25 mW and 100 kHz, respectively. This corresponds to peak laser intensity of 4.7E+09 W/cm$^2$. We use [A, B] notation to denote a particular pyramid within 5x5 array. The results reported below pertain to the pyramid/tip in a [1, 1] position. A half wave plate rotated the polarization angle of the laser with a speed of 1 degree per second. The beam images recorded at different incident laser light's polarization are shown in Fig. 3(a-d). Similarly to what we observed in our previous experiment [15], three side lobes and a bright center spot in Fig. 3(a) indicate the emission from the pyramid sides and the sharp nano-tip, respectively. The orientation of the side lobe pattern versus the pyramid indicate that the corresponding emission originates from the sides of the pyramid, not from the edges. While we observed the dependence of the side lobe pattern on the polarization of the incident light and its intensity, studies of strong-field photoemission from flat diamond surface are outside of the scope of this work. As the polarization angle of the laser light varied between 0° and 720°, the intensity of the electron beam emitted from an apex of a single pyramid varied. We integrated the intensity over the beamlet emitted from the apex to understand its polarization dependence. Polarization dependence measurements were precisely repeatable, and the observed behavior of the current was very similar to that we previously had seen in our laser-triggered photoemission study at Stanford University [15]. Figure 3(e) shows the intensity of the beam emitted from the nano-tip with a typical apex as a function of the polarization angle. The intensity of the electron beam is at its maximum when the laser electric field is aligned with the nano-tip axis.

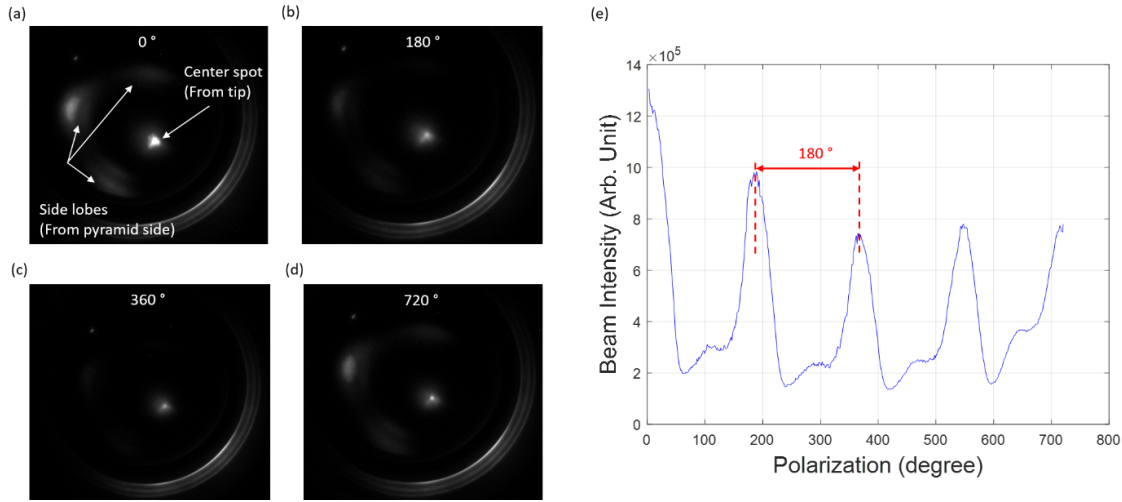

**FIGURE 3.** Emission images and collected beam intensity from the nano-tip. Photographs of polarization rotation angle with (a) 0°, (b) 180°, (c) 360°, and (d) 720°. (e) Plotted beam intensity from the nano-tip as a function of the polarization rotation angle.

## 2. Measurement of the dependence of emitted current on laser's power

As a second experiment, we measured the dependence of the current associated with a center spot in Fig. 3(a-d) on the power of the driving laser. For this measurement, we varied laser output power and repetition rate. The voltage across the MCP stack that controls the gain was adjusted in steps as needed, because the dynamic range of the digital image brightness is much less than the total range of the beam current in this experiment. The image brightness in the regions with different MCP gains was normalized sequentially with coefficients calculated by taking duplicate images at the transitions between steps, where the only changing parameter was MCP gain. Figure 4 shows the beam intensity of the center spot plotted as a function of laser peak intensity on a double logarithmic scale. The power dependence indicates a very steep onset which is typically observed in the lower optical power limit in strong-field photoemission, following the power function ($y \sim x^{8.9}$). The exponent of the power dependence suggests that nine-photon absorption process might be the primary mechanism responsible for this emission. This power dependence was similar to those measured previously by our team at Stanford University [15]. However, due to the current limitation of our setup we could not corroborate weak dependence of the exponent on the wavelength of the laser light.

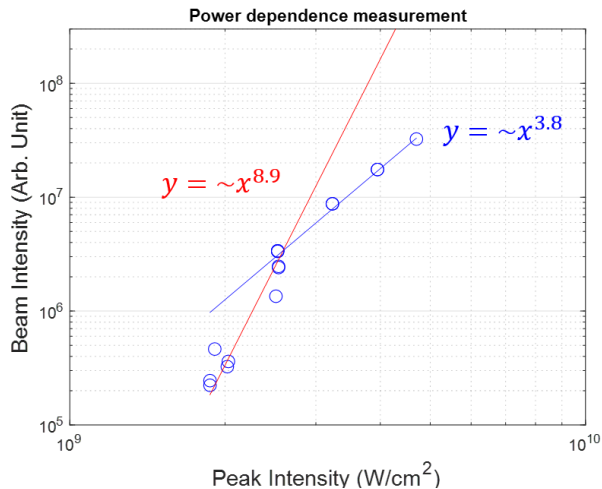

**FIGURE 4.** Electron beam intensity produced by the diamond nano-tip as a function of the laser's peak intensity.

## CONCLUSIONS

At Los Alamos we conducted experimental observation of the strong-field photoemission from a single tip of diamond pyramid. The pyramid was a part of the diamond field emission array cathode and was driven by an infrared laser with the wavelength of 1035 nm. We studied dependence of the beam emission on polarization angle of the incident laser and observed significant maximums of emission when the orientation of the electric field aligns with the axis of the pyramid tip. We also investigated the power dependences of the electron emission at the wavelength of 1035 nm. We demonstrated that as the incident laser intensity increases, the emission current emitted from the nano-tip first increases as a power function with exponent of about 9, then starts to roll over. We reason that single diamond tips can be employed as electron beam source for applications requiring intense ultra-short electron beams such as dielectric laser acceleration.

## ACKNOWLEDGMENTS

The authors gratefully acknowledge the support of Los Alamos National Laboratory (LANL) Laboratory Directed Research and Development (LDRD) program. This work was performed, in part, at the Center for Integrated Nanotechnologies, an Office of Science User Facility operated for the U.S. Department of Energy (DOE) Office of Science. Los Alamos National Laboratory, an affirmative action equal opportunity employer, is managed by Triad National Security, LLC for the U.S. Department of Energy's NNSA, under contract 89233218CNA000001.

## REFERENCES

1. R. Ganter, R. Bakker, C. Gough, S.C. Leemann, M. Paraliev, M. Pedrozzi, F. Le Pimpec, V. Schlott, L. Rivkin, A. Wrulich, Phys. Rev. Lett. **100** (2008) 2–5.
2. M. Krüger, "Attosecond physics in strong-field photoemission from metal nanotips", Ph. D thesis, Ludwig Maximilians University of Munchen, 2013.
3. W. Wu, C.K. Dass, J.R. Hendrickson, R.D. Montaño, R.E. Fischer, X. Zhang, T.H. Choudhury, J.M. Redwing, Y. Wang, and M.T. Pettes, Appl. Phys. Lett. **114**, (2019).
4. A. Tafel, S. Meier, J. Ristein, and P. Hommelhoff, Phys. Rev. Lett. **123**, 146802 (2019).
5. E.A. Peralta, K. Soong, R.J. England, E.R. Colby, Z. Wu, B. Montazeri, C. McGuinness, J. McNeur, K.J. Leedle, D. Walz, E.B. Sozer, B. Cowan, B. Schwartz, G. Travish, and R.L. Byer, Nature **503**, 91 (2013)
6. W.E. King, G.H. Campbell, A. Frank, B. Reed, J.F. Schmerge, B.J. Siwick, B.C. Stuart, and P.M. Weber, J. Appl. Phys. **97**, (2005).
7. J.D. Jarvis, H.L. Andrews, C. a. Brau, B.K. Choi, J. Davidson, W.-P. Kang, Y.- M. Wong, J. Vac. Sci. Technol. B Microelectron. Nanom. Struct. **27** (2009) 2264.
8. E. Simakov, H. Andrews, M. J. Herman, K. M. Hubbard, and E. Weis, AIP Conf. Proc. **1812**, 060010 (2017).
9. K. E. Nichols, H. L. Andrews, D. Kim, E. I. Simakov, M. Conde, D. S. Doran, G. Ha, W. Lie, J. F. Power, J. Shao, C. Whiteford, E. E. Wisniewski, S. P. Antipov, and G. Chen, Appl. Phys. Lett. **116**, 023502, (2020).
10. H. L. Andrews, K. E. Nichols, D. Kim, E.I. Simakov, S. Antipov, G. Chen, M. Conde, D. Doran, G. Ha, W. Liu, J. Power, J. Shao, and E. Wisniewski, IEEE Trans. Plasma Sci. **48**, 2671 (2020).
11. D. Kim, H.L. Andrews, B.K. Choi, and E.I. Simakov, Proc. 2018 Adv. Accel. Concepts Work. (2018).
12. J. Shao, M. Schneider, G. Chen, T. Nikhar, K.K. Kovi, L. Spentzouris, E. Wisniewski, J. Power, M. Conde, W. Liu, and S. V. Baryshev, Phys. Rev. Accel. Beams **22**, 123402 (2019).
13. D. Kim, H.L. Andrews, B.K. Choi, R.L. Fleming, C.K. Huang, T.J.T. Kwan, J.W. Lewellen, K. Nichols, V. Pavlenko, and E.I. Simakov, Nucl. Instruments Methods Phys. Res. Sect. A Accel. Spectrometers, Detect. Assoc. Equip. **953**, 163055 (2020).
14. V. Pavlenko, H.L. Andrews, R.J. Aragonez, R.L. Fleming, C. Huang, D. Kim, T.J.T. Kwan, A. Piryatinski, and E.I. Simakov, Proc. 2018 Adv. Accel. Concepts Work. (2018).
15. E. I. Simakov, H. Andrews, R. Fleming, D. Kim, V. Pavlenko, D. Black, and K. Leedle, in 10th Int. Part. Accel. Conf. (2019), pp. 2130–2133.